\documentclass[twoside,english]{paper}
\usepackage[T1]{fontenc}
\usepackage[latin9]{inputenc}
\usepackage{geometry}
\usepackage{color}
\usepackage{babel}
\usepackage{refstyle}
\usepackage{units}
\usepackage{textcomp}
\usepackage{amsmath}
\usepackage{graphicx}
\usepackage[unicode=true,pdfusetitle,
 bookmarks=true,bookmarksnumbered=false,bookmarksopen=false,
 breaklinks=false,pdfborder={0 0 0},pdfborderstyle={},backref=false,colorlinks=true]
 {hyperref}

\makeatletter

\AtBeginDocument{\providecommand\figref[1]{\ref{fig:#1}}}
\RS@ifundefined{subsecref}
  {\newref{subsec}{name = \RSsectxt}}
  {}
\RS@ifundefined{thmref}
  {\def\RSthmtxt{theorem~}\newref{thm}{name = \RSthmtxt}}
  {}
\RS@ifundefined{lemref}
  {\def\RSlemtxt{lemma~}\newref{lem}{name = \RSlemtxt}}
  {}

\makeatother

\begin{document}
\title{Evaluation protocol for revealing magnonic contrast in STXM-FMR measurements}
\author{{\large{}Benjamin Zingsem}\textsuperscript{{\large{}1,2}}{\large{},
Thomas Feggeler}\textsuperscript{{\large{}1}}{\large{}, Ralf Meckenstock}\textsuperscript{{\large{}1}}{\large{},
Taddäus Schaffers}\textsuperscript{{\large{}3}}{\large{}, Santa
Pile}\textsuperscript{{\large{}3}}{\large{},}\\
{\large{}Hendrik Ohldag}\textsuperscript{{\large{}4}}{\large{},
Michael Farle}\textsuperscript{{\large{}1}}{\large{}, Heiko Wende}\textsuperscript{{\large{}1}}{\large{},
Andreas Ney}\textsuperscript{{\large{}3}}{\large{}, Katharina Ollefs}\textsuperscript{{\large{}1}}}
\institution{\textsuperscript{1}Faculty of Physics and Center for Nanointegration
(CENIDE), University Duisburg-Essen, \\
47057 Duisburg, Germany, \\
\textsuperscript{2}Ernst Ruska Centre for Microscopy and Spectroscopy
with Electrons and Peter Grünberg \\
Institute, Forschungszentrum Jülich GmbH, 52425 Jülich, Germany \\
\textsuperscript{3}Institut für Halbleiter- und Festkörperphysik,
Johannes Kepler Universität, 4040 Linz, \\
Austria \\
\textsuperscript{4}SLAC National Accelerator Laboratory, 94025 Menlo
Park, CA, United States}
\maketitle

\paragraph*{Abstract}

We present a statistically motivated method to extract magnonic contrast
from STXM-FMR measurement with microwave frequencies of the order
of $\unit[10]{GHz}$. With this method it is possible to generate
phase and amplitude profiles with a spatial resolution of about $\unit[30]{nm}$
given by the STXM resolution, furthermore this method allows fo a
rigoros transformation to reciprocal $\vec{k}$-space, revealing $\vec{k}$-dependent
magnon properties.

\paragraph*{Introduction}

Collective oscillatory spin states, called magnons, can be excited
in magnetic materials at microwave frequencies. Their spectral characteristics
are determined by, and serve as a characterization method for, all
magnetic parameters in a magnetic system \cite{Farle1998}. Magnons
can be used as carriers of information in data processing, such as
quantum computing and spin-wave logic \cite{Khitun2010,Grundler2015,Chumak2015},
with the potential to superseed conventional electronics in many ways.
Various established techniques are employed for measuring magnonic
excitations, the most prominent ones being Ferromagnetic Resonance
(FMR)\cite{Vonsovskii1966} and Brillouin light scattering (BLS)\cite{Demokritov2001},
also neutron scattering \cite{Izyumov1970,Chatterji2006}, SEM with
polarization analyzer (SEMPA)\cite{Koike1984} and spin polarized
scanning tunneling microscopy (SP-STM)\cite{Balashov2006}, to name
a few. FMR spectroscopy is used to investigate spectral properties,
while BLS is a surface sensitive technique to measure the spatial
distribution of spinwaves down to the resolution limit of visible
light \cite{Demokritov2001}.

The technique of which we discuss the extraction of spatially resolved
of amplitude and phase of magnons is a combination of Scanning Transmission
X-Ray Microscopy (STXM) with FMR. Here the effect of X-ray Magnetic
Circular Dichroism (XMCD) is used, where the scattering of circular
polarized X-Ray with electrons in a material is dependent on the spin
of the electrons involved, i.e. the occupation of minority and majority
spin channels in the electron density of states. In the following
we discuss an evaluation technique to extract magnonic information
from such measurements. This includes spectral and spatial resolution
of magnonic egienstates in real and reciprocal space, previously not
attained.
\begin{figure}
\includegraphics[width=1\textwidth]{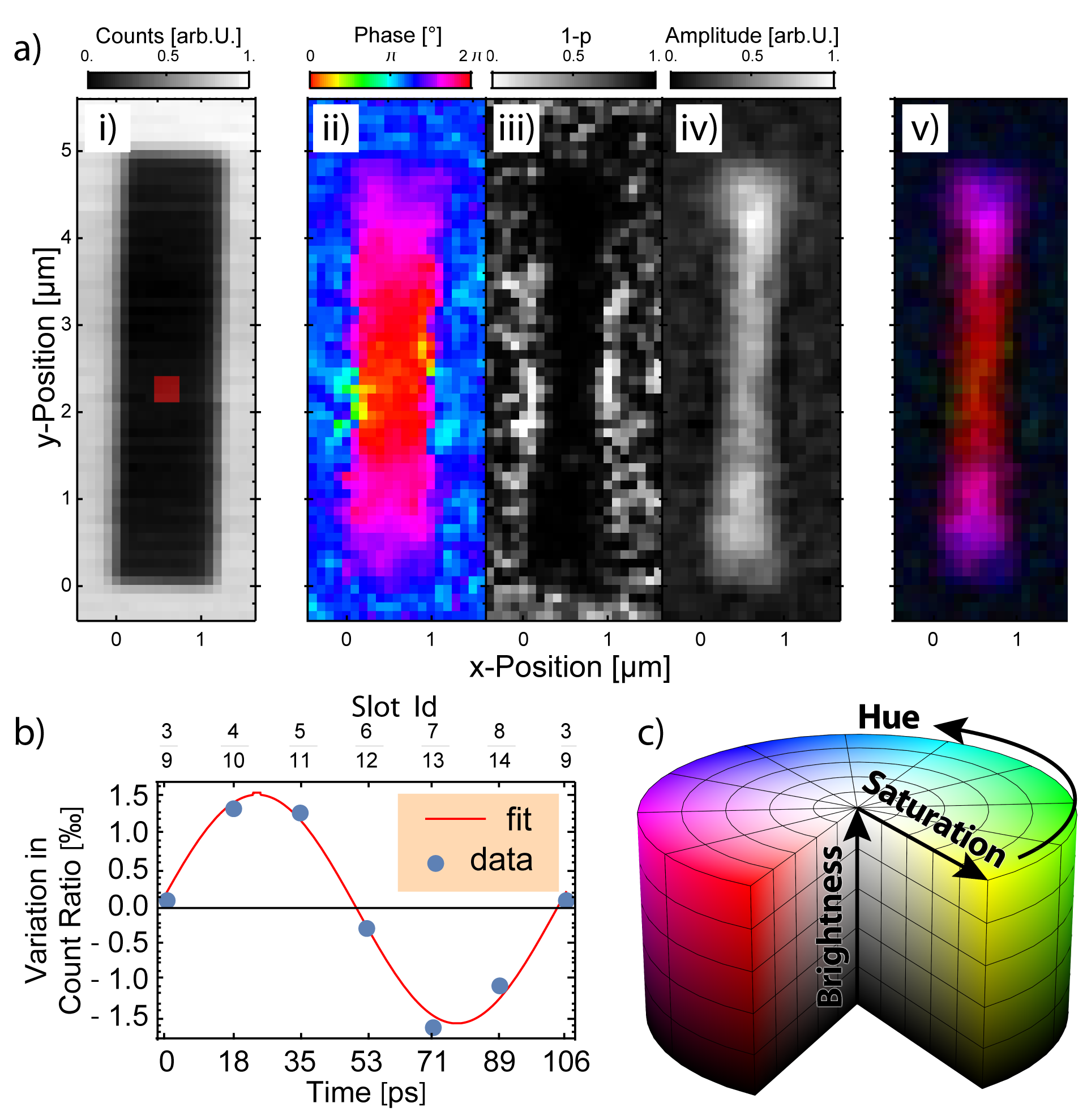}

\caption{Magnonic contrast image: a) i) Depiction of the X-Ray transmission
count rate across the sample. ii) - iv) color channels used to encode
magnonic information, where ii) depicts the phase distribution encoded
as the cyclic hue channel, iii) represents the p-value of the statistical
analysis at each pixel, encoded as the saturation and iv) represe
ts the spin-wave amplitude encoded as the bightness value in v). v)
Assembled HSB image of the magnonic excitation. b) time dependent
recording acquired in the red region in the center of a)i). c) Three
dimensional representation of the HSB color scale as used in a)v).\label{fig:Magnonic-contrast-image}}
\end{figure}
\begin{figure}
\includegraphics[width=1\textwidth]{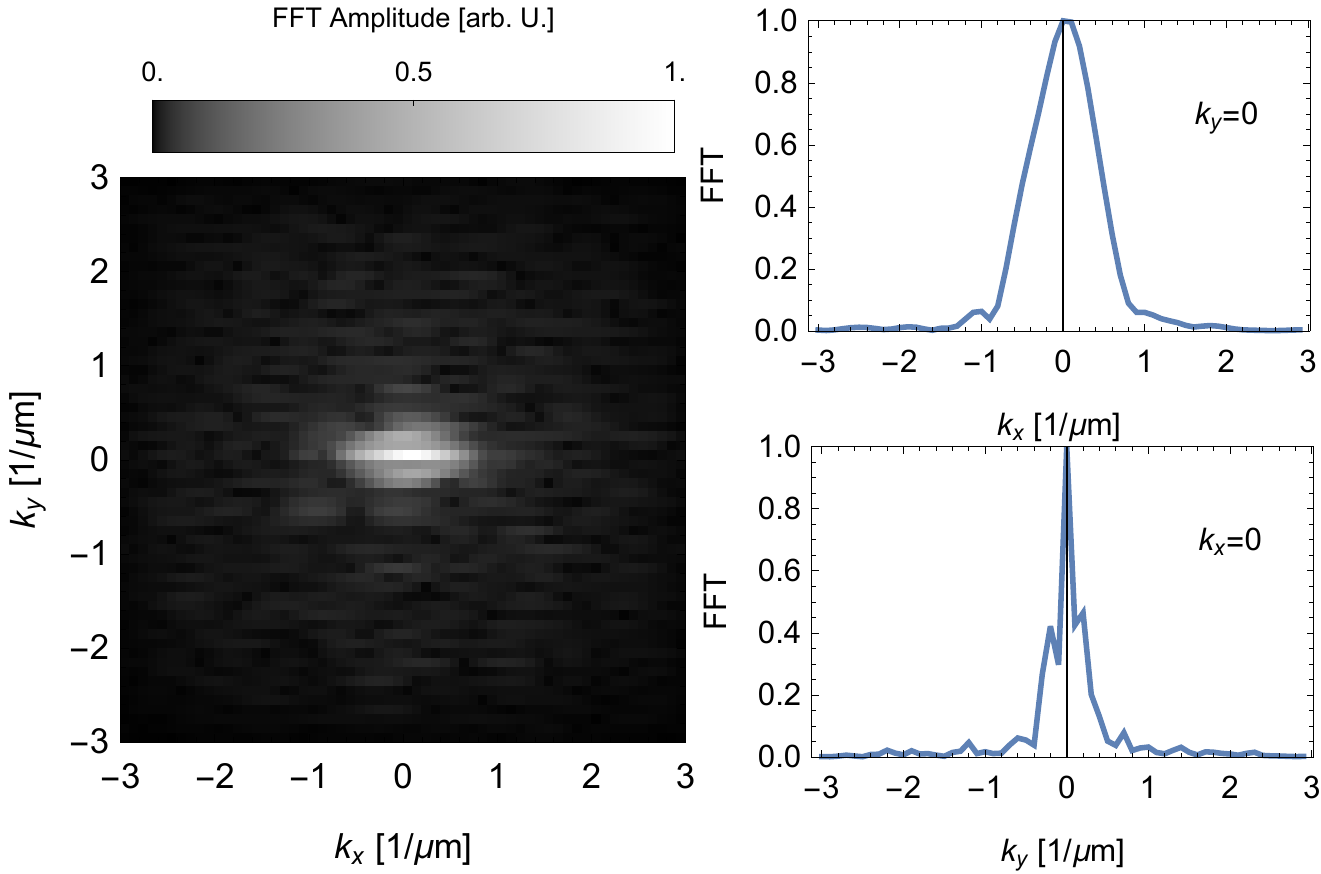}

\caption{Spatial Fourier transformation of the phase and amplitude values depicted
in Fig. \figref{Magnonic-contrast-image} a)ii) and a)iv). The $k_{x}$
dependence shows a broad distribution around $k_{x}=0$ which is most
likely dominated by the spatial confinement of the sample itself.
The $k_{y}$ distribution on the other hand shows a narrow distribution
around $k_{y}=0$ with two distinct modes visible at $k_{y}=\pm\unit[0.2]{\text{\textmu}m^{-1}}$
which demonstrates that the observed oscillation is composed of two
counter-propagating waves with a wavelength of about $\unit[5]{\text{\textmu}m}$.
\label{fig:Spatial-Fourier-transformation}}
\end{figure}

\paragraph*{Main Body}

\subparagraph*{Experimental setup and Measurement technique}

The experimental setup combining a Ferromagnetic Resonance spectrometer
and a Scanning Transmission X-Ray Microscope allows the element specific
and time resolved characterization of magnetization dynamics with
high spatial resolution \cite{Bonetti2015} of the order of $\unit[30]{nm}$.
Using X-ray magnetic circular dichroism (XMCD) a contrast proportional
to the difference in the number of minority and majority spins at
the 3d orbitals is detected. For this purpose circular polarized X-rays
are focused on the sample using a zone plate. The transmitted X-rays
are detected by an X-ray avalanche photodiode, located behind the
sample. A static magnetic field is applied perpendicular to the wavevector
of the X-rays. The sample itself is positioned in a micro resonator\cite{Narkowicz2005}
and a microwave magnetic field is applied perpendicular to the static
field and parallel the X-Ray propagation direction. The microwave
frequency is synchronized to the klystron frequency of the synchrotron
(SSRL: 476.315 MHz). In this case a microwave frequency of 9.446 GHz
is selected, which corresponds to the 20th harmonic of the klystron
frequency subtracted by 1/6, (see \cite{Bonetti2015} for details).
Using this setup we are detecting 6 points in time of a microwave
cycle with a time distance of 18 ps, each of the points is measured
with microwave on/off respectively with the same electron bunch. As
the frequency of the electron bunches (bunch length of 50 ps) in the
storage ring of the synchrotron is 1.28 MHz, a rectangular modulation
of the microwaves is implemented at this frequency using a PIN diode,
attenuating the microwaves by -35 dB. The data received from the X-ray
diode is stored in a device providing 12 \emph{slots} , the first
6 of which record the signal with microwaves turned on, whereas the
X-Ray signal with microwaves turned off ist stored in the remaining
6 slots \cite{Bonetti2015}.

The data used in this work, to demonstrate our optimized evaluation
protocol has been obtained for a sample which consists of two Py stripes
in a T-shape arrangement, spaced $\unit[2]{\text{\textmu m}}$ apart.
Each stripe has lateral dimensions of $\unit[5]{\text{\textmu m}}$
by $\unit[1]{\text{\textmu m}}$ and a thickness of $\unit[30]{\text{nm}}$.

\subparagraph*{Evaluation method}

To extract time dependent oscillations in the MW-On state, the count
rate in the MW-Off state to has to be compared to that in the MW-On
state at each position as a means of normalization. It becomes apparent
that the results will be similar regardless whether ratio or difference
of ON/OFF states is used to compare those, when considering
\begin{equation}
\frac{a_{x,y}\left(t\right)}{b_{x,y}\left(t\right)}=\frac{a_{x,y}\left(t\right)-b_{x,y}\left(t\right)}{b_{x,y}\left(t\right)}+1\label{eq:divdiff}
\end{equation}
where $a_{x,y}\left(t\right)$ is the count rate at pixel $x$, $y$
and time $t$ in the MW-On state and similarly $b_{x,y}\left(t\right)$
is the count rate in the MW-Off state. It can be motivated that the
count rate without MW $b_{x,y}\left(t\right)=b_{x,y}$ is roughly
constant in time as no spin waves are pumped. From that with eq. \ref{eq:divdiff}
the proportionality
\begin{figure}
\begin{raggedright}
\includegraphics[width=0.95\textwidth]{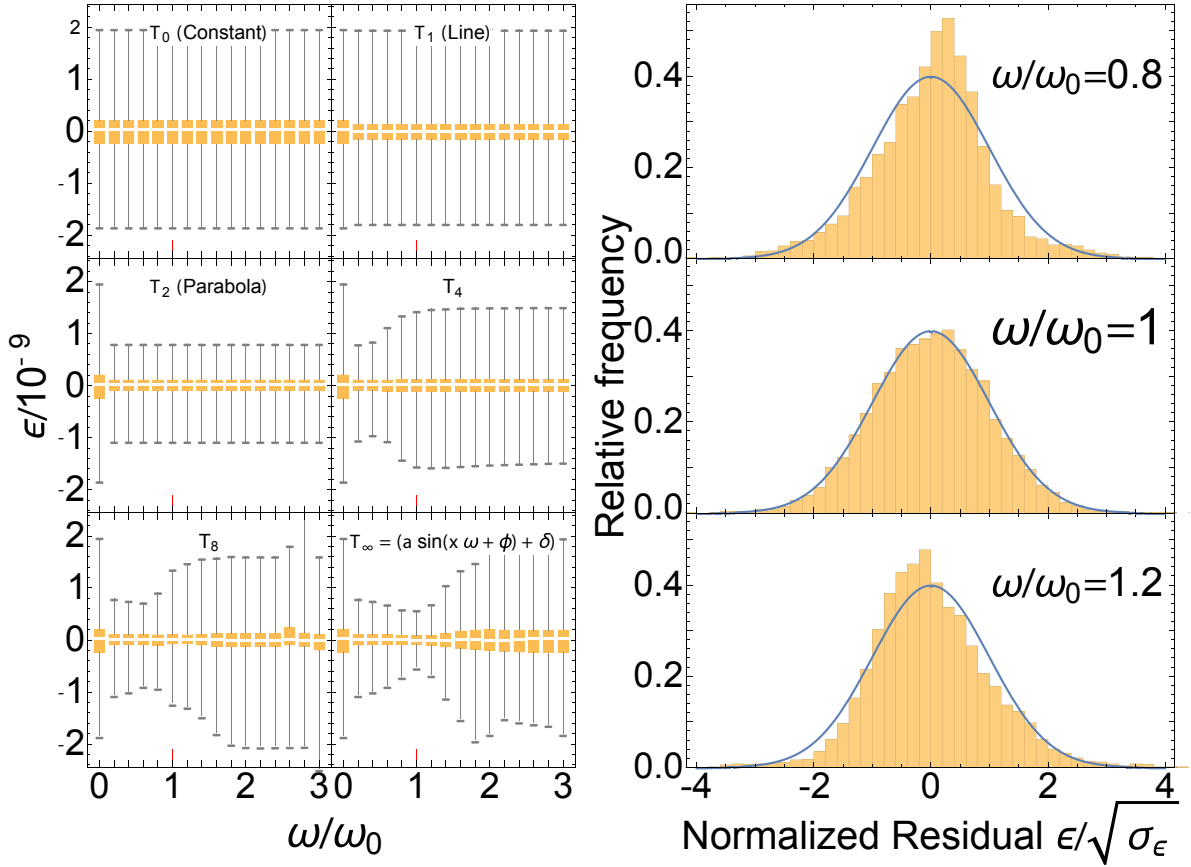}
\par\end{raggedright}
\caption{Analysis of fit residuals: In this Graphic the concatenated residuals
of the fits performed at each pixel in Fig. \ref{fig:Magnonic-contrast-image}
are analyzed. To the left, the distribution of residuals $\epsilon$
is shown in form of a Box-Chart for various fit models as a function
of the frequency. The Orange Boxes indicate the margin in which half
of the residuals are scattered, with a bright line near the center
marking the median. The gray lines mark the position of minimal and
maximal value of the residuals. To the right, the distribution of
normalized residuals for the sin model (Eq. \ref{eq:sinemodel}) is
depicted as a histogram. The normalization factor $\sqrt{\sigma_{\epsilon}}$
was calculated as the variance across the residuals. The orange bars
mark the relative count rate for each value in the set of residuals.
The blue curve is a Gaussian distribution of the form $\frac{1}{\sqrt{2\pi}}\exp\left(-\frac{x^{2}}{2}\right)$.
A clear convergence to a narrow distribution is observed for the sine
model at $\omega=\omega_{0}$. Simultaneously the shape of the distribution
of residuals near that value closely approximates a Gaussian distribution.
This would be expected if the residuals are the result of random deviations
or noise. Thus we can conclude that no additional signals are present
in the observed data and that there is no polynomial function with
three or fewer variables, that has more explanatory power than the
sine model (Eq. \ref{eq:sinemodel}).\label{fig:Analysis-of-fit-residuals}}
\end{figure}
\begin{equation}
\frac{a_{x,y}\left(t\right)}{b_{x,y}\left(t\right)}\propto a_{x,y}\left(t\right)-b_{x,y}\left(t\right)\label{eq:divdiff2}
\end{equation}
follows. The influence of noise (or signal other than the desired
magnonic contrast), however, is vastly different when comparing ratio
and difference as a means of relating the signals. This can be estimated
through the gradient of the left and right side of eq. \ref{eq:divdiff2}
In the quantities $a$ and $b$ where 
\begin{align}
\nabla_{a,b}\left(\frac{a}{b}\right) & =\left(\begin{array}{c}
\frac{1}{b}\\
-\frac{a}{b^{2}}
\end{array}\right)\\
\nabla_{a,b}\left(a-b\right) & =\left(\begin{array}{c}
-b\\
a
\end{array}\right).
\end{align}
Hence, fluctuations in $b$ have a non-linear effect on fluctuations
in the ratio, whereas fluctuations in $a$ and $b$ have a linear
effect on fluctuations in the difference. Thus, the signal to noise
ratio should be considered when deciding how to compare MW-On and
MW-Off signal. Except for the perceived noise level, these operations
yield equivalent results in the evaluation of magnonic contrast and
both can be used in qunatitative analysis as one can be transformed
into the other using eq. \ref{eq:divdiff}.

A $\sin$ like magnonic response to the $\sin$ like microwave excitation
is expected, such that the magnonic contrast can now be extracted
by fitting a $\sin$-function to the time dependent data at each point
on the sample that was normalized using one of the aforementioned
methods. Also it is expected that the dynamic response of the sample
oscillates at the same frequency as the microwave driving it. Figure
\hyperref[fig:Magnonic-contrast-image]{1 b)} depicts a sine wave
fitted into the time dependent signal at one point on the sample.
The model is given as

\begin{equation}
A\sin\left(2\pi\frac{\omega}{\omega_{0}}t+\phi\right)+d\label{eq:sinemodel}
\end{equation}
where $A$ accounts for an amplitude of the oscillation, $\omega_{0}$
is the excitation frequency, $t$ is the time in periods, $\phi$
accounts for a phase shift relative to $t=0$ and $d$ describes an
offset, which is usually close to $0$ for the case of $a-b$ and
close to $1$ when analyzing $\frac{a}{b}$. Note that only $A$,
$\phi$ and $d$ are free parameters in the fit, and the physical
assumption is that $\omega=\omega_{0}$.

Performing this fit at each pixel, we can now extract the spatial
distribution of the fit parameters, i.e. the spatial distribution
of amplitude and phase. Additionally a simple measure for the validity
of this fit can be given by the statistical p-value. It measures the
likelihood that the data-set to which the fit was applied originates
from a random distribution rather than a distribution following the
assumed model. The p-value therefore gives a measure for how adequate
our fit is at each pixel, where small p-values indicate a good fit.
These three values can then be encoded in an image as seen in Fig.
\hyperref[fig:Magnonic-contrast-image]{1 a)v)}. Here the \emph{hue,
saturation, brightness} (HSB) color scale was used, where the phase
was encoded into the \emph{hue} channel, the amplitude was encoded
into the \emph{brightness} channel, and one minus the p-value ($1-p$)
was encoded into the saturation. Thus bright pixels correspond to
high amplitudes, while highly saturated pixels correspond to a very
sine-like oscillation. One can quickly spot, that encoding this information
into an image like this immediately reveals the spatial structure
of the excited magnonic eigenstate, including phase information. The
next task it to investigate the $\vec{k}$-vectors this state comprises.
This can be achieved, by introducing a complex value $z_{x,y}$ at
each pixel $x,y$, where $\left\Vert z_{x,y}\right\Vert =A$ and $\arg\left(z_{x,y}\right)=\phi$.
Applying a Fourier transformation in the spatial coordinates $x$
and $y$ then transforms the spatial wave to reciprocal space, revealing
the $\vec{k}$-vectors involved in this state. Figure \ref{fig:Spatial-Fourier-transformation}
illustrates the $\vec{k}$ resolution for the spatial eigenstate observed
in \ref{fig:Magnonic-contrast-image}. The broad distribution around$\vec{k}_{x}=0$
can be attributed to the spatial confinement of the sample, while
the narrow distribution around $\vec{k}_{y}=0$ shows distinct maxima
for the involved $\vec{k}$-vectors assembling the eigenmode along
the long axis of the stripe.

In the previous section it was physically motivated, to assume a sine-function
as representative for the investigated oscillations, one could argue,
that this motivation is weak, as there may, for example, be multiple
sine functions (i.e. Fourier components) involved in one oscillation,
or that the data does not represent a sine function at all, but rather
a simpler function. A simple test can now be performed by assuming
different values for $\omega$. Figure \ref{fig:Analysis-of-fit-residuals}
shows the distribution of fit residuals under variation of $\omega$,
starting from $\omega=0$ to $\omega=2\omega_{0}$ in steps of $0.1\omega_{0}$.
It is immediately apparent, that the residuals converge to a narrow
Gaussian distribution at $\omega=\omega_{0}$ confirming the assumed
frequency. Further still, the model itself can be motivated, by trying
simpler functions as models for the fit. For this purpose, we considered
a Taylor expansion of the sine function. This yields a set of polynomials,
starting from $0^{\mathrm{th}}$ order and progressively approaching
the sine function where $\mathrm{T_{n}}$ denotes a Tailor series
expansion to the nth order. The residuals as a function of the frequency
for various polynomials in this series are depicted in Fig. \ref{fig:Analysis-of-fit-residuals}.
It is quickly noted, that the convergence is best in only one case,
namely when using the sine-function with an assumed frequency of $\omega=\omega_{0}$.
In this case the residuals closely approximate a Gaussian distribution
indicating that they are randomly distributed and not the result of
a systematic deviation. 

\paragraph*{Summary}

We have demonstrated, how magnonic contrast can be extracted from
STXM-FMR measurements with statistical rigor even in the case of few
time steps. Not only frequency and spatial phase-and-amplitude profiles
can be extracted, but a transposition to reciprocal space can be performed
to identify the prevailing $\vec{k}$-vectors in a collective magnonic
eigenstate. Furthermore, we have shown that our method is -- beyond
its physical motivation -- statistically motivated, independent of
physical assumptions. It carries statistical significance and allows
quantitative deductions about the magnonic properties.

\bibliographystyle{vancouver}
\bibliography{STXM_FMR_Auswerung}

\end{document}